\documentclass[times,10pt,twocolumn]{article} 
\usepackage{latex8}
\usepackage{times}

\usepackage{url}
\usepackage{graphicx}
\usepackage{moreverb}

\newcommand{\US}{U.S.\@}
\newcommand{\etal}{et al.\@}
\newcommand{\myFig}[1]{Figure~\ref{#1}}
\newcommand{\cfFig}[1]{cf.\@{} \myFig{#1}}

\newcommand{\mySec}[1]{Section~\ref{#1}}

\newcommand{\myquote}[2]{
  \hfill\parbox{3in}{\small
      {``}{\em #1\/}{''}\\
      -- #2\\
  }
}

\newcommand{\mySection}[1]{\Section{#1}}

\begin{document}

\title{
  Research Challenges in Management and Compliance of Policies on the Web
}

\author{
  Holger M. Kienle$^*$\\
  University of Victoria\\
  Victoria, Canada\\
  {\tt hkienle@acm.org}
  \and
  Hausi A. M\"uller\\
  University of Victoria\\
  Victoria, Canada\\
  {\tt hausi@cs.uvic.ca}
}

\date{}

\maketitle

\thispagestyle{empty}
\pagestyle{empty}

\begin{abstract}
  In this paper we argue that policies are an increasing concern for
  organizations that are operating a web site. Examples of policies
  that are relevant in the domain of the web address issues such as
  privacy of personal data, accessibility for the disabled, user
  conduct, e-commerce, and intellectual property.
  Web site policies---and the overarching concept of web site
  governance---are cross-cutting concerns that have to be addressed
  and implemented at different levels (e.g., policy documents, legal
  statements, business processes, contracts, auditing, and software
  systems). For web sites, policies are also reflected in the legal
  statements that the web site posts, and in the behavior and features
  that the web site offers to its users.  Both policies and software
  tend to evolve independently, but at the same time they both have to
  be kept in sync. This is a practical challenge for operators of web
  sites that is poorly addressed right now and is, we believe, a
  promising avenue for future research.
  In this paper, we discuss various challenges that policy poses for
  web sites with an emphasis on privacy and data protection and
  identify open issues for future research.
\end{abstract}

\noindent \textbf{Keywords:} Internet, legal factors, compliance control

\mySection{Introduction}\label{sec:Intro}

\myquote{There is hardly a government in the world that does not have
  some form of policy about the World Wide Web.}{Diffily
  \cite{Diffily:06}}

In this paper, we want to raise awareness that policy issues for web
sites have to be addressed during the site's whole life cycle,
including requirements, design, development and maintenance. Policy
compliance and management are parts of web site governance.
Web sites are challenging from the perspectives of governance and
policy, because they operate in a highly homogeneous environment
(e.g., multiple languages, corporate entities, and jurisdictions) with
diverse users.  There are increasingly complex sites (web 2.0 sites
and sites of e-tailers) that offer functionality that rivals
shrink-wrapped consumer products---and site complexity translates to
policy complexity.  Furthermore, there may be a number of different
stakeholders involved that have different goals and objectives for the
site.
One can distinguish three important groups of stakeholders: the web
site operator, the web site's users, and lawmakers.

Understanding and addressing policy issues and requirements from the
beginning has a number of potential benefits for the development of
web sites. Policy requirements for the web site's domain can be
elicited during requirements engineering, thus making them a part of
the entire software life cycle.  This way, issues such as privacy and
security that may impact the system's architecture and design can be
addressed early on. Thus, addressing policy issues from the start can
prevent costly changes in subsequent development or maintenance
activities.

Policies and software have to be kept in sync. A policy can evolve for
various reasons (e.g., changing business needs, lobbying by
stakeholders, or legal developments) and there needs to be process and
tool support to trace policy changes down to the code.  Conversely,
changes in the code can (unwittingly) implement behavior that
contradicts a policy. Hence, there should be static and dynamic checks
to verify policy compliance and to detect policy violations.

It should not be taken lightly by an organization if its web site
violates a policy. Generally policies such as privacy and terms of use
statements can be seen as binding contracts between the operator of
the site and its users.
If the policy touches on users' rights, a violation may result in a
loss of reputation and trust, causing users to abandon the site. For
example, studies show that users are often concerned about their
privacy \cite{BGS:CACM:05}, and as a result may be outraged about a
violation of the site's privacy policy.
If the policy is mandated by law, violations can cause lengthy and
costly legal actions, severe penalties, and temporary shut-down of the
site.
Thus, risk mitigation demands that operators of web sites have to
re-evaluate constantly the conformance of the behavior of their site
with respect to various policies.

The rest of the paper is organized as follows. 
In \mySec{sec:ITG} we start out with a primer on information
technology governance in general and web site governance in particular
because both topics are closely related to policy management. In
\mySec{sec:Issues} we give an overview of the diversity of policy
issues that web sites have to address.
In \mySec{sec:Privacy} we focus on privacy policies as an example to
expose issues that need be addressed in policy management and
compliance for web sites. Privacy policies are a opportune example
because organizations are free to define their own policies, but they
are also constrained by national laws and consideration of users'
trust.  Furthermore, privacy needs to be addressed for internal
operation, but it also needs to be communicated to users of the web
site.
Based on the discussion of privacy policies, we discuss in
\mySec{sec:Complexity} the complexity of policy management for
different kinds of web sites (brochure-ware, e-commerce, and web 2.0),
and describe in \mySec{sec:Implications} implications and research
challenges that organizations have to tackle for policy management and
compliance.  \mySec{sec:Concl} closes with conclusions.

\mySection{IT and Web Site Governance}\label{sec:ITG}

\myquote{Compliance with regulations regarding internal controls,
  financial reporting, and privacy is now a substantial catalyst for
  companies to understand and invest in addressing information risk
  challenges.}{Ernst \& Young Global Information Security Survey
  \cite{EYSurvey:06}}

Policy management and compliance of software systems are increasingly
important activities that need to be addressed by most organizations.
One of the trends identified by the Ernst \& Young Global Information
Security Survey is that ``the impact of compliance continues to grow''
\cite{EYSurvey:06}.

Policy management and compliance is typically an integral part of
\emph{information technology (IT) governance}, which is driven by the
realization that an organization's software systems are the hub around
which its business activities revolve. Consequently, an organization's
IT capabilities can no longer be treated as a black box by the
companies stakeholders \cite{WikipediaITG}. Instead, various
stakeholders across the organization have to coordinate and resolve
policy issues. One key challenge for stakeholders is ``balancing
compliance at all costs, compared to compliance at an affordable
cost'' \cite{PCWelt:05}.

Web sites are part of an organization's IT infrastructure and as such
have to be incorporated into IT governance. Furthermore, web sites are
an organization's interface to the public. Many organizations and
businesses use web sites to post information for, and to interact with
consumers. The term \emph{web site governance} \cite{Diffily:06} has
been suggested to emphasize the importance of web sites and to stress
that web sites pose particular challenges for governance.
In fact, policy compliance of a web site (e.g., a banking portal) is
equally important to a (backend) software system (e.g., a financial
transactions system). Furthermore, there may be complex interactions
between them.

Depending on the nature of an organization, governance can be more
lightweight or heavyweight. Diffily believes that web site governance
``does not require a huge and unwieldy bureaucracy, just some plainly
written guidelines and clear executive oversight''
\cite[p.~288]{Diffily:06}.
Diffily suggests to have a web site management team (WMT) that is in
charge of the web site and (1) defines the organizations web strategy,
(2) sets the site's high level goals and ensures the achievement of
the goals, and (3) monitors overall performance. The WMT is also
responsible to ensure that processes for site management are in place
and that these processes address policy issues adequately.
In order to enable the WMT to operate effectively, there should be
dedicated tool support that assists in the tasks of policy management
and compliance checking. These tools should probably operate at
different levels of granularity, including high-level dashboards
\cite{Marcus:interactions:06} and lower-level information such as
policy goals and ontologies.

\mySection{Policy and Legal Issues of Web Sites}\label{sec:Issues}

\myquote{No longer an information `wild, wild, west,' the Internet
  increasingly is influenced by legal considerations.}{Baker in
  \cite{Isenberg:02}}

Policy issues for web sites touch on many diverse areas. An
organization may have a security policy, privacy policy, corporate
identity policy, ethics policy, customer care policy, etc. 

Often policy issues interact with laws or regulations. For example, an
increasing number of countries have data protection laws that web
sites have to adhere to; however, even if a country has no such law it
may still make good sense for a web site to address this issue in
their privacy policy to increase consumer trust. Another cross-cutting
issue is jurisdiction because web sites are accessible by users
worldwide.
Depending on the nature of the site (e.g., financial or healthcare),
domain-specific policy and legal issue may emerge. An extreme example
is a gambling web site, because it targets a domain that is heavily
regulated by most states and countries.

Examples of typical policy issues for web sites that are influenced by
legal considerations are \cite{Diffily:06} \cite{KGTM:IJBIS:08}
\cite{Isenberg:02} \cite{Takach:03}:
\begin{description}
\item[criminal damage:] A web site may (inadvertently) cause harm to a
  site's user. A site may host a virus (e.g., in a web page's
  JavaScript or a downloadable file) that deletes data on the user's
  computer. In this case, the site operator may be liable for
  negligence.
  To give another example, courts have applied \emph{trespass law} to
  prohibit frequent, unwanted spidering of a web site if a meaningful
  harm to the site's computer system (e.g., resource drain) could be
  shown \cite{Samuelson:CACM:03}.

\item[freedom of expression:] Certain kinds of content on web sites
  raise the issue of free speech and its restrictions (e.g., product
  defamation, slander and libel, company secrets, and hateful ideas).
  Generally, it is often difficult to decide if free speech applies or
  not. For example, is unwanted but harmless email protected by free
  speech, or could such email be prohibited based on trespass law?
  Since the freedom of expression varies significantly by
  jurisdiction, the decision where to operate a site may be
  significant.

\item[intellectual property:] The content, design, functionality, and
  domain name of a web site may be protected by intellectual property
  (i.e., copyright, trademarks, and patents). On the one hand, a web
  site site has to protect its own intellectual property; on the other
  hand, the site has to ensure that it does not violate the
  intellectual property rights of third parties.
  In the context of copyright (and freedom of expression) it is
  significant that posting material to a web site is considered an act
  of publishing.

\item[electronic commerce:] E-commerce refers to the trading of goods
  and services over the Internet. In response to the large volume of
  commerce on the Web, legislation has introduced rules to govern
  online transactions. As a result, agreements made via the Internet
  and electronic signatures are legally binding, consumer rights
  protection applies to goods purchased on the web, purchases on the
  web are taxable, etc.

\item[accessibility for the disabled:] A web site should be accessible
  to users with disabilities. The \US{} has amended the Rehabilitation
  Act to require Federal agencies to make their web site accessible
  (Section 508). As a result, persons with disabilities may file
  administrative complaints or bring civil actions in Federal court
  against agencies that fail to comply with the requirements of
  Section 508.
  There are commercial web sites that have chosen to address
  accessibility in their web sites. For example, General Electric
  posts an explicit statement on its site that states its current
  accessibility features (\url{www.ge.com/accessibility.html}).

\end{description}

The above issues are meant to illustrate the broad range of policies
that interact with legal requirements. There are certainly other
policy considerations that a web site needs to address. Another
important topic is privacy and data protection, which is discussed in
detail in \mySec{sec:Privacy}.

\mySection{Data Protection and Privacy Policies}\label{sec:Privacy}

\myquote{Traditionally, policy specification has not been an explicit
  part of the software development process. This isolation of policy
  specification from software development often results in policies
  that are not in compliance with system requirements and/or
  organizational security and privacy policies, leaving the system
  vulnerable to data breaches.}{He \etal{} \cite{HOAJ:IWIA:06}}

Privacy can be defined as ``the ability of an individual or group to
seclude information about themselves and thereby reveal themselves
selectively'' \cite{WikipediaPrivacy}. In the context of information
technology, an important concern is data protection of digitally
stored information (e.g., health, criminal, financial, genetic,
ethnic, and location information). Importantly, more low-level data
should be also considered private such as stated user preferences
(e.g., language setting) and interactions with the system (e.g.,
executed search engine queries, and visited web pages and sites).

In response to growing privacy concerns, many countries have passed
laws that govern the treatment of sensitive data. As any software
system, web sites have to meet legal obligations.  The European Union
has enacted Directive 95/46/EC in 1995 \cite{EuDirective},
which requires organizations that collect personal data to register
with the government and to take precautions against data misuse.
Furthermore, organizations have to inform individuals about the
reasons for collecting information about them, to provide access to
the data and to correct wrong data.
This directive has to be implemented by states that are members of the
EU. For example, the UK has the Data Protection Act (1998), and
Germany has the {\em Bundesdatenschutzgesetz}
(2001).\footnote{\url{http://ec.europa.eu/justice_home/fsj/privacy/law/implementation_en.htm}}
Examples of non-EU countries with data protection laws are Canada (2004),
Australia (2001), Japan (2005), and Switzerland (1993).
In the \US{} there is no single act or law that addresses privacy.
Instead, there are different laws that touch on privacy and data
protection. For example, the Health Insurance Portability and
Accountability Act (HIPAA) establishes regulations for the use and
disclosure of health information, the Gramm-Leach-Bliley Act regulates
the privacy of customers of financial institutions, and the Children's
Online Privacy Protection Act of 1998 (COPPA) addresses privacy issues
for children under 13.

What constitutes private data is not easy to decide. Knight and
Fitzsimons give the following example: ``Some people regard receiving
a flood of 'junk mail' as an invasion of privacy, others regard it as
just a part of modern living, or even a chance to be informed''
\cite{KF:90}. An important, but unresolved, question in law is who
owns the (private) data of users. According to Taipale, ``a
fundamental issue, as yet not fully resolved to everyone's
satisfaction in the context of emerging technologies, is whether data
{\em about} an individual (whether disclosed by that individual or
otherwise obtained) should `belong' to that individual in any kind of
sense that would invoke legal mechanisms of ongoing control---i.e.,
some notion of property---or perhaps even a renewal of `expectations'
of privacy for secondary uses---after it [is] shared or otherwise
becomes known'' \cite[p.~154]{Taipale:YJOLT:04}. For example, if a web
site collects information about a user (e.g., search queries), does
the user own that data or the collecting entity?

There are potential legal concerns whenever a web site collects,
processes and stores data that contains information of or about its
users. In the following, we briefly give examples of web site features
that are interacting with privacy policies.
\begin{description}
\item[logging of data:] The log files of a web site may store
  information about the interactions of a user with the system.
  Weitzner \etal{} advocate \emph{policy-aware transaction logs} that
  are responsible for ``recording information-use events that may be
  relevant to the assessment of accountability to some set of
  policies'' \cite{WABFH:CACM:08}.
  There is often a clash between privacy protection on the one hand,
  and security and auditing concerns on the other.\footnote{The
    tension between auditing requirements and privacy for log files
    becomes apparent by the following pun: ``If logs mention private
    information they are forbidden and if they do not, they are
    useless'' \cite{EMY:TrustBus:07}.} On the one hand, there may be
  legal requirements that make it necessary that data about users is
  retained for a certain period of time. On the other hand, logged
  data may have to be anonymized in order to protect the privacy of
  users.
  For privacy protection it is important whether the logs guarantee
  certain properties such as {\em complete anonymity}
  \cite{EMY:TrustBus:07}.

\item[profiling and personalization:] Many commercial web sites
  collect data of user interactions to personalize pages
  \cite{Kobsa:CACM:07} \cite{Volokh:CACM:00}. For example, Amazon
  generates recommendations of books that are personalized by the
  user's own history of viewing and buying books as well as the buying
  habits of other ``similar'' users. Google is working on personalized
  searches that take a person's search history into account
  \cite{Sullivan:SEL:07}; the user's search history is kept in the
  Google Web History. For such services, web sites have to ensure that
  the site's privacy policy is communicated to users, that the site
  does not expose private data to other users, and that users have a
  certain control over the collected information (e.g., correction of
  wrong information).

\item[distribution and transmission of data:] Users of a web site will
  potentially access it from all over the world. However, certain
  countries have policies that govern the transmission of personal
  data across borders.
  The EU privacy directive mentioned above prohibits the exporting of
  personal data to a country that does not provide an adequate level
  of privacy protection. So far the EU has recognized few countries,
  among them Canada and
  Switzerland.\footnote{\url{http://ec.europa.eu/justice_home/fsj/privacy/thridcountries/index_en.htm}}
  For countries not recognized by the EU, the user would be required
  to explicitly give consent to the web site operator.

\item[communication of policies to the user:] Last but not least, web
  sites have to post their privacy policy. These policies are
  important for users to understand how their data will be treated by
  the site operator. Ant\'on \etal{} say, ``often, the only guide
  users have as to how an institution will use, disclose, and store
  sensitive information is via its online privacy policies. Thus,
  users should expect these privacy policy documents to accurately
  describe an institution's privacy practices in a clear and
  easy-to-understand manner'' \cite{AEVJG:IEEESP:07}.
  The following is an excerpt a legal notice from the web site of a
  large American corporation in
  1998:\footnote{\url{http://web.archive.org/web/19980530081620/www.valero.com/html/legal_notice.htm}}
  \begin{quote}
    ``Any visitor to the Valero web site who provides information to
    Valero agrees that Valero has unlimited rights to such information
    as provided, and that Valero may use such information in any way
    Valero chooses. Such information as provided by the visitor shall
    be non-confidential.''
  \end{quote}
  Even though an organization is free to fashion its own policy, in
  practice there are many legal and ethical constraints that need to
  be taken into account. For example, web sites of healthcare
  providers have to reflect the legal requirements of HIPAA in their
  privacy policies \cite{AEVJG:IEEESP:07}. It seems rather unlikely
  that the above legal notice would be considered as adequate
  nowadays.
\end{description}
There are probably many more policy issues that need to be addressed
besides the examples given above.
However, these examples are sufficient to expose the complexity of
managing privacy policies for web sites.

Furthermore, the interactions of policies and web sites increase with
the complexity of the site. This issue is discussed in more detail in
\mySec{sec:Complexity}.

\mySection{Policy Complexity of Web Sites}\label{sec:Complexity}

Policy issues are typically more pronounced for more complex web site
is. For discussion, we define three kinds of sites with increasing
sophistication in terms of functionality and user interactions:
\begin{description}
\item[brochure-ware:] These sites provide information that users can
  browse (e.g., to obtain information about products and services that
  they can obtain off-line) \cite{TH:ICSE:01}. User do not have to log
  on to the site and the site is static in the sense that it looks the
  same for all users.
\item[e-commerce:] These sites are run by companies that sell products
  online. They may be pure online retailers (\emph{e-tailers}) or have
  a \emph{clicks-and-bricks} hybrid business model \cite{PG:CACM:03}.
  To place orders, users have to create an account.
\item[web 2.0:] These sites are characterized by sophisticated
  functionality that often rival shrink-wrapped software products
  (e.g., Google GMail\footnote{\url{http://mail.google.com}} and Adobe
  Photoshop
  Express\footnote{\url{https://www.photoshop.com/express/landing.html}}).
  These sites typically offer a participatory and interactive user
  experience \cite{CK:FirstMonday:08}, which is typically realized
  with technologies such as AJAX, mashups, blogs, Wikis, and RSS
  \cite{Murugesan:ITPro:07}.
\end{description}
The above classification is an idealization because concrete web sites
typically have features that blur into other groups. For example, a
brochure-ware site may have a form or questionnaire that users can fill
out to provide feedback to the site operator, and e-commerce sites
often have some kind of personalization (e.g., Amazon's wishlists) or
user-generated content (e.g., book reviews of users).

The simplicity of brochure-ware sites makes policy management and
compliance comparably easy to accomplish. Since all content is
supplied and maintained by sources within the organization, processes
can be defined that mandate compliance checks by a central authority.
For example Diffily suggests that ``all ideas for new information or
applications must be approved by a WMT before work commences. \dots In
order to get a development approved, a proposal must be submitted and
a collective decision made about whether it is suitable for the site
(perhaps based on advice from the editor).''
\cite[p.~294]{Diffily:06}. As a result, tracking and assessment of
content-related issues such as intellectual property, accessibility,
and freedom of expression can be implemented in a straightforward
manner.
Since a brochure-ware site can be realized with static HTML, many
security threats are mitigated or do not exists (e.g., script-based
and SQL injection attacks). Furthermore, since brochure-ware sites do
not ask for personal information from users, there are only privacy
issues of tracking the movements of users on the site.

Policy issues of e-commerce sites are significantly more difficult to
handle than those for brochure-ware sites. The web site itself is more
complex because it needs functionality to manage user accounts and
purchasing. As a consequence, e-commerce sites have to manage personal
data of users such as address and billing information. Furthermore,
they store---permanently or temporarily---sensitive information about
users such as purchased items and defaults in payments. As a
consequence, the site has to adhere to data protection laws of various
jurisdictions. For example, organizations may be required to report
stolen data to affected people. Following California in 2002, many
states in the U.S. have adopted security breach notification
laws.\footnote{\url{http://www.ncsl.org/programs/lis/cip/priv/breachlaws.htm}}
These laws require an organization to notify all residents within the
law's state if it believes that personal, non-public information of
residents has been stolen.
The EU is considering a similar law---amending EU Directives
2002/22/EC and 2002/58/EC---that would force telecommunications
companies to tell customers when personal data security has been
breached.\footnote{\url{http://www.out-law.com/page-8741}}
Since e-commerce sites do business with users, they have to adhere to
consumer protection laws, handle taxation, and deal with fraudulent
transactions. Even though complexity increases, it appears that many
policy aspects can still be managed by a central authority.

Web 2.0 sites add further complexity for policy management and
compliance due to several characteristics. These sites have typically
user-generated content where users are \emph{conducers}, that is, they
``both consume creative works and simultaneously add creative content
to those same works'' \cite{Reuveni:SSRN:08}. As a result content is
no longer created exclusively within an organization and as such
policy compliance of content is difficult to enforce.  Specifically,
web sites have to address liability issues for users' misuse of
intellectual property (e.g., copyright infringements), misinformation,
slanderous comments or other harmful data.
Web 2.0 sites are realized by using web browsers as thin clients where
most of the application logic and data storage is performed by distant
Internet servers. This approach is discussed as cloud computing
\cite{Hayes:CACM:08} and Platform-as-a-Service (PaaS)
\cite{Lawton:IEEEC:08}. Hayes points out that cloud computing ``raises
awkward questions about control and ownership: If you move to a
competing service provider, can you take your data with you? Could you
lose access to your documents if you fail to pay your bill?  Do you
have the power to expunge documents that are no longer wanted?''
\cite{Hayes:CACM:08}.

From a privacy perspective, the service provider has to establish
polices on how to manage and protect personal information entrusted by
its users. In contrast to e-commerce sites, web 2.0 may collect
personal data on a much larger scale. This is especially the case for
consolidated information service providers such as Google that
accumulate huge amounts of personal data depending on the extent of
services used by a particular user (e.g., various kinds of searches,
emails, appointments, contacts, documents, news alerts, and
financials) \cite{Conti:NSPW:06}.
Web sites that store personal data also have to be prepared on whether
and how they defend their users' privacy rights.
Hayes describes the following scenario: ``a government agency presents
a subpoena or search warrant to the third party that has possession of
your data. If you had retained physical custody, you might still have
been compelled to surrender the information, but at least you would
have been able to decide for yourself whether or not to contest the
order.  The third-party service is presumably less likely to go to
court on your behalf'' \cite{Hayes:CACM:08}.  In fact, the Department
of Justice in the \US{} ordered Google in 2005 to hand over two month
of search queries. Google refused---primarily on the grounds of trade
secrets, but also because of privacy concern---and a court decided in
Google's
favor.\footnote{\url{http://googleblog.blogspot.com/search/label/privacy}}

From the users' perspective, published policies of web sites are
important because they spell out the users' rights and obligations.
Examples of such statements are a site's privacy policy and terms of
use. With increasing complexity of the web site, these policies also
increase in complexity.
While brochure-ware sites can be satisfied by covering only general
issues (e.g., license to use, disclaimer, linking, and intellectual
property), e-commerce sites also have to address issues such as order
acceptance, pricing information, exporting of goods, and disclaimers
for special goods such as medicines. Web 2.0 sites are also more
complex than brochure-ware because they have to address complex user
interactions with the site involving personal data.

To summarize, web sites have to define policies that cover a diverse
range of issues. Increasing complexity in a web site typically
translates into increasing complexity of policy management. Also, a
subset of a web site's policies has to be communicated to users in the
form of statements posted on the site.

\mySection{Implications and Research Challenges}\label{sec:Implications}

The implementation and management of policy issues is a major
challenge for an organization. Many of the policies have to be
reflected in the organization's web site and kept in synch with
internal policies and the web site's functionality.  Also, policies
can be complex to define and maintain.
For example, Ant\'on \etal{} have analyzed the Privacy Rule of HIPAA,
formalizing its content as restricted natural language statements
\cite{BVA:RE:06}. They found 46 rights and 80 obligations that need to
be addressed. Presumably, most of these rights and obligations apply
to the web site's content and application logic. In another study,
Ant\'on \etal{} show that privacy policies of web sites have indeed
changed significantly after HIPAA came into effect
\cite{AEVJG:IEEESP:07}.
Ant\'on \etal{} also have evaluated web site policies in the financial
sector; they concluded that ``compliance with the existing legislation
and standards is, at best, questionable'' \cite{AEHSB:IEEESP:04}.

\begin{figure*}[htb]
  \begin{center}
    \fbox{
      \includegraphics[width=.68\textwidth]{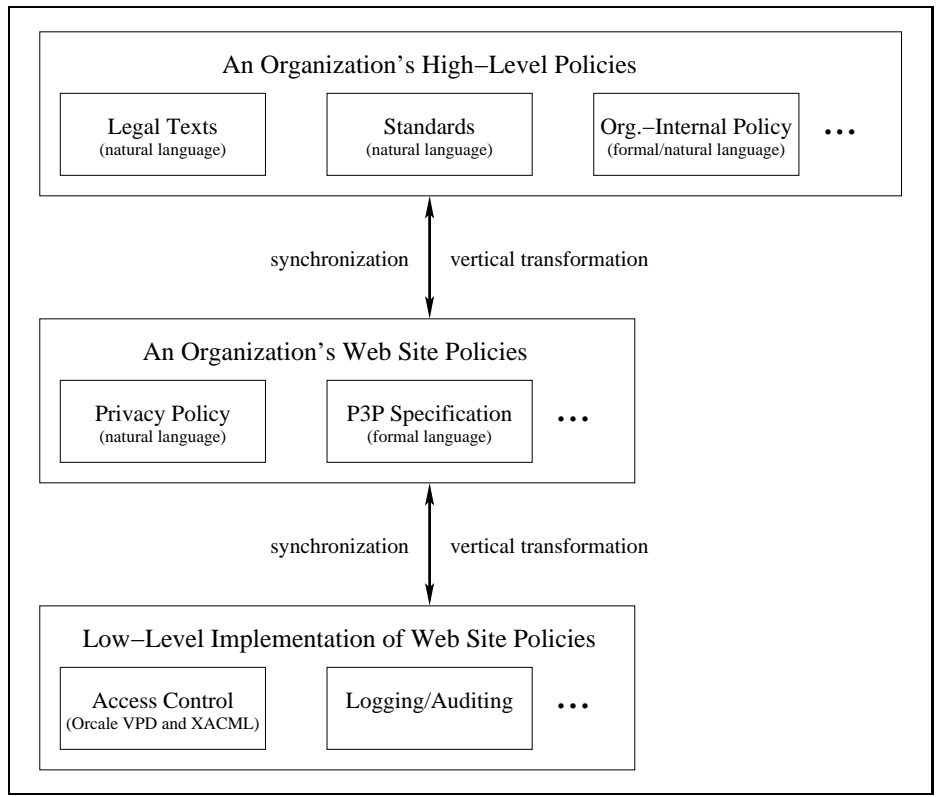}
    }
    \caption{Policies at different levels of abstractions and transformations of these policies}
    \label{fig:ptrans}
  \end{center}
\end{figure*}

Policies have to be managed at different levels of abstraction and in
different representations. \myFig{fig:ptrans} groups policies into
three tiers: high-level policies that apply to an organization as a
whole, policies that are specific to the organization's web site(s),
and the implementation of these policies.
Ant\'on \etal{} introduce a framework for online privacy policies that
is also based on three tiers; they distinguish the top tier
(principles of privacy practices), middle tier (security policies),
and bottom tier (enforcement in the physical layer)
\cite{ABLY:CACM:07}.

At the top tier in \myFig{fig:ptrans} there are documents that define
policies at a high level of abstraction and in natural language.
Examples of such policies are legal texts (e.g., Acts enacted by the
\US{} Congress), standards (e.g., W3C accessibility guidelines), and
internal policies of the organization. The latter can be in natural
language or a dedicated policy language.\footnote{For example,
  Google's organizational goals with respect to privacy are outlined
  in its Code of Conduct available at
  \url{http://investor.google.com/conduct.html#1}:
  \begin{quote}
    ``As we develop great products that serve our users' needs, always
    remember that we are asking users to trust us with their personal
    information. Preserving that trust requires that each of us respect
    and protect the privacy of that information.''
  \end{quote}}
Such high-level policies have to be substantiated into web site
policies that are posted on the site.\footnote{For example, Google has
  a general Privacy Policy (about 1900 words) that is augmented with
  service-specific privacy policies. For AdWords, Google claims that
  its ``conversion tracking server complies with P3P privacy
  policies''
  (\url{http://adwords.google.com/support/bin/answer.py?hl=en&answer=6358}).}
These policies are expressed in natural language so that users can
read them, but also in formal languages for automated processing.
Transforming a legal text into a formal language such as a goal model
can be (partially) automated (e.g., \cite{KZBAC:ER:08}). An example of
a dedicated formal language is the W3C's Platform for Privacy
Preferences (P3P), which is an XML-based format that enables web sites
to encode their data collection and data-use practices.
Finally, policies have to be encoded in the web site's content and
functionality. For example, an operator that decides not to collect
personal information from children (in response to COPPA) may decide
to post notices and to implement safeguards in their web site that
communicate and enforce this particular policy. 
More generally, policies concerning data protection of personal
information that are communicated to users, have to be faithfully
reflected in the site's behavior. This means for instance that the
part of a database that stores private information has to be
safeguarded with access control mechanisms (e.g., Oracle supports
row-level security with its virtual private database feature
\cite{Nanda:OracleMag:04}). Similarly, the application logic needs
access control mechanisms (e.g., implemented with OASIS's eXtensible
Access Control Markup Language (XACML) or IBM's Enterprise Privacy
Authorization Language (EPAL) \cite{Anderson:SWS:06}).
Furthermore, logging and auditing data that contains private
information must not violate the privacy policy (e.g., by anonymizing
the data).

If policy issues are considered for web sites, they affect their whole
life cycle and other cross-cutting concerns:
\begin{description}

\item[requirements analysis and policy representation:] Policies have
  to be treated as first-class entities in requirements engineering.
  Furthermore, they have to be formalized so that it is possible to
  analyze and reason about them. This may be accomplished with general
  formalisms such as goal models \cite{Lamsweerde:ISRE:01} or
  domain-specific representations such as P3P and EPAL. Policies have
  characteristics of both quality attributes (e.g., security) and
  functional requirements (e.g., data retention for a certain amount
  of time). Also, since policies are often fuzzy and ambiguous it must
  be possible to model degrees of uncertainty.
  
  Currently, approaches such as P3P are too limited to describe
  fine-grained privacy rules \cite{ABLY:CACM:07}. It is an important
  research challenge to come up with a suitable representation, which
  may be customized from a general approach or developed bottom-up
  specifically for the web.

\item[policies and design models:] It is quite likely that policies
  are impacting the design of web sites. A trivial example would be a
  requirement that the privacy policy is accessible (via a hyperlink)
  from each page of the site. A more complex example is the impact of
  users' privacy settings on the data visible to other users.  Policy
  requirements should be reflected as constraints and other
  annotations in the web site's application, navigation, and
  presentation model. For example, depending on privacy settings, the
  navigation model may enable or disable navigation paths to
  particular sensitive content, and the presentation model may replace
  one widget with another (e.g., a name alias instead of the user's
  real name with his or her picture).
  In order to represent policies during design, web-design
  methodologies such as WebML \cite{CFM:WWW:01} or OOHDM
  \cite{SR:TAPOS:98} have to be suitably augmented.

\item[testing for compliance:] Web sites have to be tested for policy
  compliance before they are deployed. In fact, there are many
  examples of privacy violations caused by wrong implementations of
  privacy features. For example, in Facebook supposedly private
  annotations were made visible to all users
  \cite{Goodin:Register:07}.
  Testing for policy compliance is complicated by the fact that web
  sites have to support diverse user and environmental profiles,
  resulting in highly varied behaviors at the client side
  \cite{PTG:WSE:07}.

  Generally, testing of user interactions of web sites exhibits
  similar problems than GUI testing \cite{MNX:JSME:05}. It is an open
  research problem how to effectively specify test cases for policy
  compliance and how to (automatically) evolve these tests if the web
  site evolves.

\item[monitoring for compliance:] In addition to off-line testing, web
  sites should be monitored at run-time for compliance. A large
  corporate presence may be composed of multiple web sites that are
  administered by different entities in different countries. As a
  result, such sites may resemble systems of systems or even
  ultra-large scale systems in some respects \cite{Pollak:SEI:06}. For
  such systems, ``policies will have to reconcile diverse and
  competing objectives while providing complete and unambiguous
  semantic content sufficiently to govern distributed system
  development, evolution, and operation''
  \cite[p.~119]{Pollak:SEI:06}. For web sites that cannot be centrally
  controlled and tested, run-time monitoring becomes increasingly
  important. A simple example would be a crawler that periodically
  checks whether accessibility guidelines, linking policies, and
  content do conform to company policies and legal requirements.

  Ideas from autonomic computing could be adapted to monitor web sites
  in terms of their data, operations, and communication
  \cite{Muller:WCRE:06}. Furthermore, if a failure occurs in the web
  site (e.g., because low-level access control has detected a privacy
  policy violation), strategies for reconfiguring and self-healing
  could be employed (e.g., to gracefully degrade the site's
  functionality, or to achieve the desired functionality without
  violating a particular access control path).

\item[policy transformations:] Policies have to be represented at
  different levels of abstraction. As a result, it is necessary to
  implement \emph{vertical} transformation of policies between the
  three tiers identified in \myFig{fig:ptrans}.
  For example, general privacy policies of the organization that are
  stated in natural language in the top tier have to be reflected in
  the middle tier (e.g., the web site's privacy policy, its privacy
  seal, and P3P specification) as well as the low-level implementation
  (e.g., database and client/server-side code).

  There are also \emph{horizontal} transformations (e.g., translation
  of a policy encoded in a formal-language representation into a
  human-readable web policy in the middle tier).
  It is an open question how to transform policy requirements into
  policy constraints.  Furthermore, policy constraints should be
  automatically translated into testing code and run-time checks.

\item[policy synchronization and traceability:] Because policies are
  encoded in different forms at different levels of abstraction
  (\cfFig{fig:ptrans}), they have to be re-synchronized if one policy
  changes. Synchronization is facilitated if policies are formalized
  in machine-readable representations and if they can be automatically
  transformed. However, natural language policies may not be fully
  automated. Furthermore, seemingly unrelated changes in the behavior
  and operations of the web site may contradict certain policy rules
  as an undesirable side-effect. For example, a low-level data
  replication features (that is transparent to upper layers) may be in
  violation of privacy law if the replication involves data
  transmissions that cross borders (e.g., to a data center that is
  located in another country).

  To effectively synchronize polices and to handle errors, policies
  have to be traceable. For example, it should be possible to trace a
  high-level natural language policy to its formal representation
  (e.g., part of a goal model) and further down to the code.
  Conversely, if a fault occurs because a low-level policy check is
  violated, traceability should facilitate the maintainers to
  rationalize which higher-level policies are responsible for this
  particular check.

\item[policy negotiation:] If a web site requests a service of another
  entity (e.g., in a service-oriented setting) there needs to be some
  form of policy matching and negotiation. For example, the requested
  service has to guarantee (e.g., in a service level agreement) that
  personal data is processes in accordance with the requesting web
  site's policies.  Policy negotiation may be fully automated (e.g.,
  communicating web services) or human-involved. In the latter case,
  dedicated tool support is needed that facilitates negotiation.
  SPARCLE is an example of such a privacy management tool
  \cite{BKKF:SOUPS:05}. The W3C has developed A P3P Preference
  Exchange Language (APPEL),
  \footnote{\url{http://www.w3.org/TR/P3P-preferences/}} which allows
  users to describe privacy policies that are acceptable to them and
  to check whether a given web site accommodates them.

  Since different users and services may have different policy
  requirements, a web site has to be prepared to accommodate different
  policy settings. In this scenario, different privacy preferences of
  users may result in different behavior of the web site.
\end{description}

Another important concern besides the above challenges is how to make
existing web sites more policy-aware. For simpler web sites, it seems
feasible to reimplement them, but this may not be an option for
complex e-commerce and web 2.0 sites. When migrating a web site,
policy-aware features can be injected incrementally. For example, the
code of the web site can be reverse engineered to extract stakeholder
goal models and to establish traceability between the existing code
and the goal model \cite{YWMLL:RE:05}. The goal model can then be
augmented with policy-aware requirements and the code evolved to
reflect the changes in the goal model.

\mySection{Conclusions}\label{sec:Concl}

\myquote{Security is an emergent property of a system, not a feature.
  \dots{}Because security is not a feature, it can't be bolted on
  after other software features are codified, nor can it be patched in
  after attacks have occurred in the field.}{Hope \etal{}
  \cite{HMA:IEEESP:04}}

Hope \etal{} in the above quote stress that security is not a set of
immovable features \cite{HMA:IEEESP:04}. The same is true for policy
compliance, which is a moving target because of various forces that an
organization is confronted with. These forces can be internal ones such
as changes in strategy that affect policy as well as external ones
such as new regulations.  Given that policies will evolve, there have
to be mechanisms in place to enable a policy-aware evolution of the
site. Policy-aware evolution has to be addressed in the whole life
cycle of the web site starting with requirements.

In this paper, we have discussed policy management and compliance for
web sites, giving examples of concrete policy and legal issues.
We have then explored one particular policy issue in more detail,
namely privacy and data protection. This issue exposes the complexity
of policy management: web sites have to address it in logging of data;
profiling and personalization; distribution and transmission of data;
and in communicating their privacy policy to the users of the site.
Furthermore, policy issues are more pronounced with increasing
complexity of the site.

We believe that the policy issues that we have identified show that
web sites can no longer be developed in a vacuum without consideration
of policies and legal constraints. Treating policy requirements as
first-class entities is justified by the potentially severe adverse
consequences of ignoring them.
This paper has exposed a number of policy challenges, which can serve
as a starting point in formulating a research agenda to advance the
current state of policy management and compliance for web sites.

\section*{Acknowledgments}

Thanks to Crina Vasiliu for proofreading and commenting on an earlier
draft of this paper.

This work has been supported by the Natural Sciences and Engineering
Research Council of Canada (NSERC), the Consortium for Software
Engineering (CSER), and the Centre for Advanced Studies (CAS), IBM
Canada Ltd.

\bibliographystyle{latex8}
\bibliography{books,papers,theses,own,short}

\noindent \includegraphics[scale=.6]{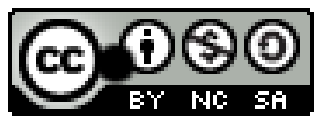}
\begin{minipage}[b]{.75\linewidth}
  {\tiny This work is licensed under a Creative Commons
    Attribution-Noncommercial-Share Alike 3.0 United States License.
    The license is available here:
    \url{http://creativecommons.org/licenses/by-nc-sa/3.0/us/}.}
  \baselineskip8pt
\end{minipage}

\end{document}